\documentclass[aps,twocolumn,a4paper,citesort]{revtex4}
\usepackage{graphicx}
\usepackage{bm}
\usepackage{latexsym}
\usepackage{psfig}
\usepackage{epsfig}

\def\x{{\mbox{\boldmath$x$}}}

\def\eps{{\epsilon}}
\def\be{\begin{equation}}
\def\ee{\end{equation}}
\def\bea{\begin{eqnarray}}
\def\eea{\end{eqnarray}}

\def\lp{\left(}
\def\rp{\right)}
\def\pp{\bm{\partial}}
\def\uu{\bm{u}}
\def\kk{\bm{k}}
\def\xx{\bm{x}}

\def\nn{\bm{\nabla}}

\begin{document}
\bibliographystyle{prsty_withtitle}

\title{Rayleigh and Prandtl number scaling in the bulk of
  Rayleigh-B\'enard turbulence}
\author{Enrico Calzavarini$^{1,2}$} 
\author{Detlef Lohse$^{3}$}
\author{Federico Toschi$^{2,4}$} 
\author{Raffaele Tripiccione$^{1,2}$}

\affiliation{$^{1}$Dipartimento di Fisica, Universit\`a di Ferrara,
  Via Paradiso 12, I-43100 Ferrara, Italy.}  

\affiliation{$^{2}$INFN, Via Paradiso 12, I-43100 Ferrara, Italy.}

\affiliation{$^{3}$Department of Applied Physics and J. M. Burgers
  Centre for Fluid Dynamics, University of Twente, 7500 AE Enschede,
  The Netherlands.}  

\affiliation{$^{4}$ IAC-CNR, Istituto per le Applicazioni del
  Calcolo, Viale del Policlinico 137, I-00161 Roma, Italy.}

\begin{abstract}
  The $Ra$ and $Pr$ number scaling of the Nusselt number $Nu$, the
  Reynolds number $Re$, the temperature fluctuations, and the kinetic
  and thermal dissipation rates is studied for (numerical) homogeneous
  Rayleigh-B\'enard turbulence, i.e., Rayleigh-B\'enard turbulence
  with periodic boundary conditions in all directions and a volume
  forcing of the temperature field by a mean gradient. This system
  serves as model system for the bulk of Rayleigh-B\'enard flow and
  therefore as model for the so called ``ultimate regime of thermal
  convection''. With respect to the $Ra$ dependence of $Nu$ and $Re$
  we confirm our earlier results \cite{loh03} which are consistent
  with the Kraichnan theory \cite{kra62} and the Grossmann-Lohse (GL)
  theory \cite{gro00,gro01,gro02,gro04}, which both predict $Nu \sim
  Ra^{1/2}$ and $Re \sim Ra^{1/2}$.  However the $Pr$ dependence
  within these two theories is  different. Here we show
  that the numerical data are consistent with the GL theory $Nu \sim
  Pr^{1/2}$, $Re \sim Pr^{-1/2}$. For the thermal and kinetic
  dissipation rates we find $\eps_\theta/(\kappa \Delta^{2}L^{-2})
  \sim (Re Pr)^{0.87}$ and $\eps_u/(\nu^3 L^{-4}) \sim Re^{2.77}$,
  also both consistent with the GL theory, whereas the temperature
  fluctuations do not depend on $Ra$ and $Pr$.
Finally, the dynamics of the heat transport is studied and put into
the context of a recent theoretical finding by Doering {\it et al.}
\cite{doe05}.
\end{abstract}
\date{\today}
\maketitle
\section{Introduction}
The scaling of large Rayleigh number (Ra) Rayleigh-B\'enard convection
has attracted tremendous attention in the last two decades
\cite{cas89,wu90,wu92,zoc90,bel93,kad01,cha97,cha01,roc01,roc02,sig94,nie00,nie01,nie03,gro00,gro01,gro02,gro03,gro04,xu00,ahl01,nik03,ver99,ver03b,ver03,cam98,sha03,du98,she96,du00,qiu01a,cio97,cil99,xia97,xia02,lam02,xi04,day01,day02,chi97,chi01a,chi02,tak96,nae97,seg98,som99,gla99,ken02,doe96,bre04}.
There is increasing agreement that in general there are no clean
scaling laws for $Nu(Ra,Pr)$ and $Re(Ra,Pr)$, apart from asymptotic
cases.  One of these asymptotic cases has been doped the ``ultimate
state of thermal convection'' \cite{kra62}, where the heat flux
becomes independent of the kinetic viscosity $\nu$ and the thermal
diffusivity $\kappa$. The physics of this regime is that the thermal
and kinetic boundary layers have broken down or do not play a role any
more for the heat flux and the flow is bulk dominated. The original
scaling laws suggested for this regime are \cite{kra62}

\bea
\label{eq1}
Nu &\sim& Ra^{1/2}(\log  Ra)^{-3/2}\ Pr^{1/2} \\
Re &\sim& Ra^{1/2}(\log Ra)^{-1/2}\ Pr^{-1/2}, \label{eq1_}
\eea 
for $Pr<0.15$, while for $0.15<Pr \lesssim 1$: 
\bea
\label{eq2}
Nu &\sim& Ra^{1/2}(\log  Ra)^{-3/2}\ Pr^{-1/4} \\
Re &\sim& Ra^{1/2}(\log Ra)^{-1/2}\ Pr^{-3/4}.
\label{eq2_}
\eea 

The GL theory also gives such an asymptotic regime which is bulk
dominated and where the plumes do not play a role \cite{gro04}
(regimes IV$_l$ and IV$_l^\prime$ of Refs.\ 
\cite{gro00,gro01,gro02,gro04}). It has the same Ra dependence as in
eqs.\ (\ref{eq1})-(\ref{eq2}), but different Pr dependence, namely

\be
Nu \sim Ra^{1/2} Pr^{1/2},
\label{eq3}
\ee
\be
Re \sim Ra^{1/2} Pr^{-1/2}.
\label{eq4}
\ee 

As a model of the ultimate regime we had suggested
\cite{loh03} homogeneous RB turbulence, i.e., RB turbulence with
periodic boundary conditions in all directions and a volume forcing of
the temperature field by a mean gradient \cite{bif03}, 

\be {{\partial \theta} \over {\partial t}} + \lp\uu\cdot \pp\rp \theta
= \kappa \partial^2 \theta + {\Delta \over L} u_z.
\label{eq5}
\ee 

Here $\theta = T + (\Delta/L) z$ is the deviation of the temperature
from the linear temperature profile $-(\Delta/L) z$.  The velocity
field $\uu(\x , t)$ obeys the standard Boussinesq equation,

\be {{\partial \uu} \over {\partial t}} + \lp \uu \cdot \pp \rp \uu =
-\nn p +\nu \partial^2 \uu +\beta g {\hat z} \theta .
\label{eq6}
\ee 

Here, $\beta$ is the thermal expansion coefficient, $g$ gravity, $p$
the pressure, and $\theta (\x , t)$ and $u_i (\x, t)$ are temperature
and velocity field, respectively. Indeed, in Ref.\ \cite{loh03} we
showed that the numerical results from eqs.\ (\ref{eq5}) and
(\ref{eq6}) are consistent with the suggested
\cite{kra62,gro00,gro01,gro02,gro04} $Ra$ dependence of $Nu$ and $Re$,
$Nu \sim Ra^{1/2}$ and $Re \sim Ra^{1/2}$. However, the $Pr$
dependences of $Nu$ and $Re$, for which the predictions of Kraichnan
\cite{kra62} and GL \cite{gro00,gro01,gro02,gro04} are different, has
not yet been tested for homogeneous turbulence: this is the first aim
of this paper (Section III). Section II contains details of the
numerics. In Section IV we study the bulk scaling laws for the thermal
and kinetic dissipation rates and compare them with the GL theory. In
that Section we study the temperature fluctuations $\theta' = \left<
  \theta^2 \right>^{1/2}$.  The dynamics of the flow, including
$Nu(t)$ and its PDF (probability density function), is studied in Section V and put into the context
of a recent analytical finding by Doering and coworkers \cite{doe05}.
Section VI contains our conclusions.

\section{Details of the numerics}
Our numerical simulation is based on a Lattice Boltzman Equation (LBE)
algorithm on a cubic $240^3$ grid. The same scheme and resolution has
already been used in \cite{cal02,bif03}.  We run two sets of
simulations in statistically stationary conditions.  The first at
fixed $Pr = 1$ varying the $Ra$ number between $9.6\cdot10^4$ and
$1.4\cdot 10^7$.  The second at fixed $Ra = 1.4\cdot 10^7$. This, the
highest value we can reach at the present resolution, was studied for
five different $Pr$ numbers, $1/10$, $1/3$, $1$, $3$ and $4$.  We
recorded shortly-spaced time series of $Nu$ and root mean squared
(rms) values of temperature and velocity and we stored a collection of
the whole field configurations, with a coarse time-spacing.  The
length of each different run ranges between $64$ and $166$ eddy turn
over times.  Our simulation was performed on a APE$mille$ machine in a
$128$ processor configuration \cite{ape99}, \cite{ape02}. Each eddy
turnover times requires on average 4 hours of computation. The total
computational time required for the whole set simulations is roughly
150 days. The total number of stored configurations is around 2000.

\section{$Nu(Ra,Pr)$ and $Re(Ra,Pr)$}
The Nusselt number is defined as the dimensionless heat flux

\bea 
Nu &=& {1\over \kappa \Delta L^{-1} } \left( \left< u_3 T
  \right>_{A,t} (z) - \kappa \left<  \partial_3 T \right>_{A, t}
  (z) \right)
  \nonumber 
\\ 
  &= & {
  \left< u_3 \theta
  \right>_{A,t} (z) \over \kappa \Delta L^{-1} } - 1
\label{eq7}
\eea 
where the average $\left< ... \right>_{A,t} $ is over a horizontal
plane and over time. From eqs.\ (\ref{eq5})-(\ref{eq7}) one can
derive two exact relations for the volume averaged thermal dissipation
rate $\eps_\theta = \kappa \left< (\partial_i \theta )^2\right>_V$ and
the volume averaged kinetic dissipation rate $\eps_u = \nu \left<
  (\partial_i u_j )^2\right>_V$, namely 
\be \eps_u = {\nu^3\over L^4}
Nu Ra Pr^{-2} ,
\label{eq8}
\ee
\be
\eps_\theta = \kappa {\Delta^2 \over L^2} Nu.
\label{eq9}
\ee 

One can therefore numerically compute $Nu$ in three different ways:
(i) from its direct definition (\ref{eq7}), (ii) from the volume
averaged kinetic dissipation rate (\ref{eq8}), (iii) from the volume
averaged thermal dissipation rate (\ref{eq9}).

\begin{figure}[!th]
\begin{center}
\setlength{\unitlength}{1.0cm}
\begin{picture}(1,1)
\put(-3.5,-0.5){(a)}
\end{picture}
\includegraphics[width=8.cm]{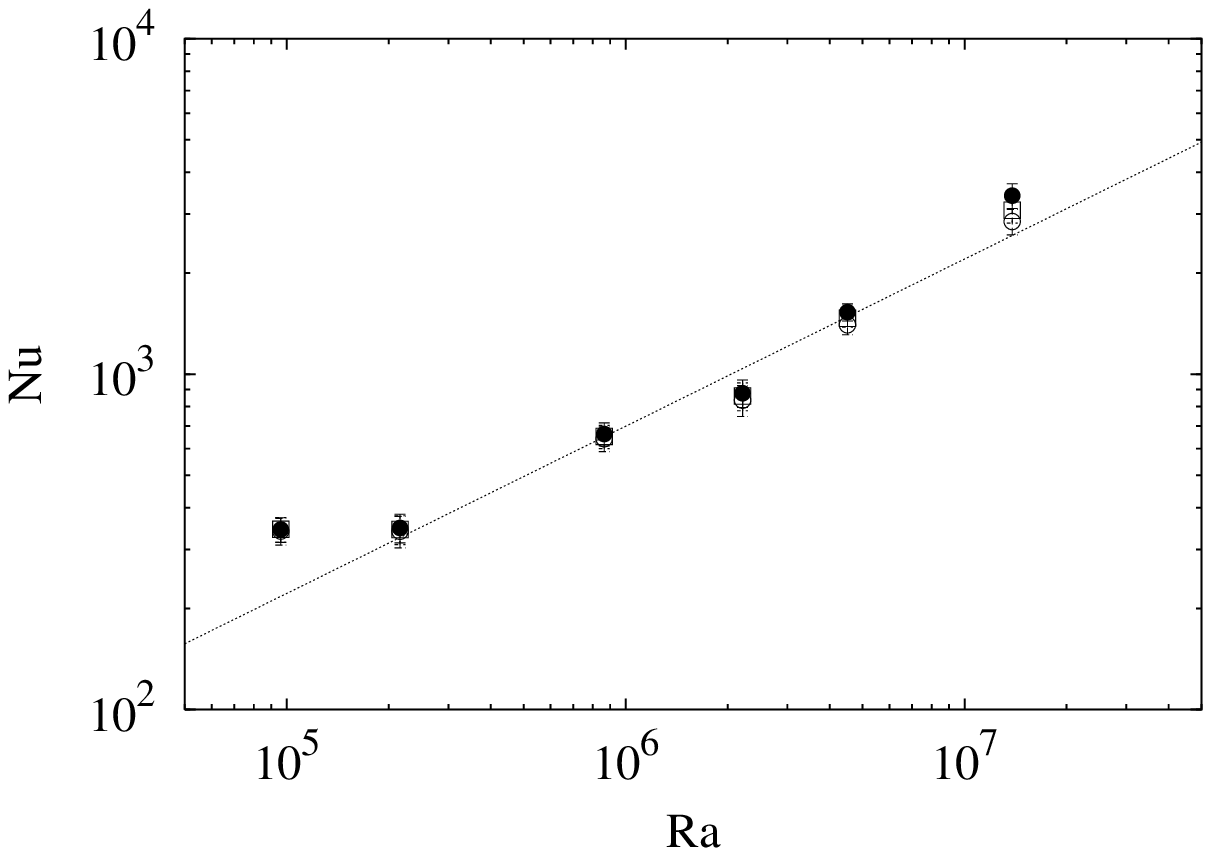}
\setlength{\unitlength}{1.0cm}
\begin{picture}(1,1)
\put(-3.5,-0.5){(b)}
\end{picture}
\includegraphics[width=8.cm]{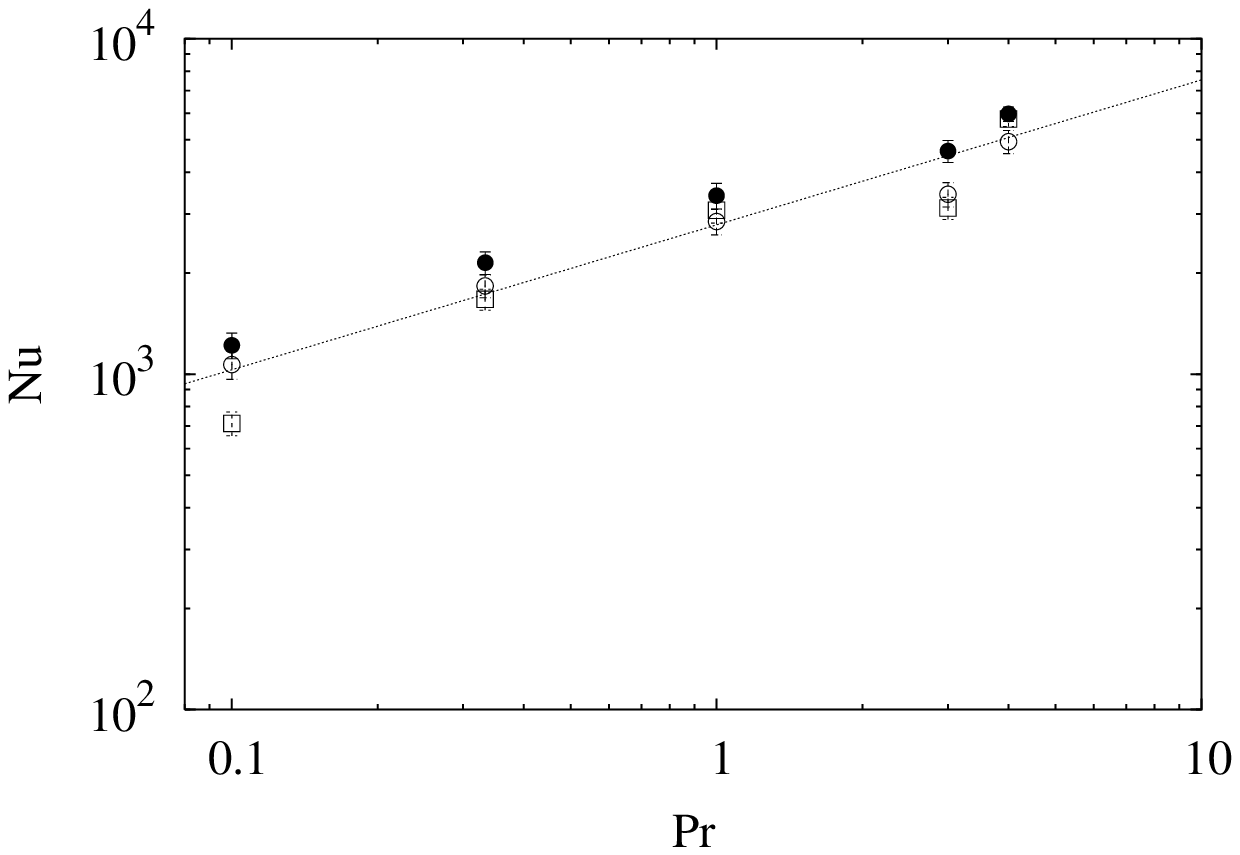}
\caption{
  (a) $Nu(Ra)$ for $Pr=1$, computed in three different ways:
  ($\bullet$) using Eqn. (\ref{eq7}), ($\Box$) using Eqn. (\ref{eq8}),
  and ($\circ$) from Eqn. (\ref{eq9}). The power law fits, performed
  on the mean value of the three different estimates and for $Ra >
  10^5$, gives a slope $0.50 \pm 0.05$.  (b) $Nu(Pr)$ for $Ra= 1.4
  \cdot 10^7$, fit performed as before, with a resulting slope of
  $0.43 \pm 0.07$.
\label{fig1}
}
\end{center}
\end{figure}

The results are shown in Figure \ref{fig1}a as a function of $Ra$ for
$Pr=1$. There is very good agreement of $Nu$ obtained from the three
different methods for all $Ra$, giving us further confidence in the
convergence of the numerics. If we fit all data points beyond
$Ra=10^5$ with an effective power law, we obtain $Ra\sim Ra^{0.50 \pm
  0.05}$, consistent with the asymptotically expected law $Nu \sim
Ra^{1/2}$ \footnote{In our previous paper \cite{loh03} the overall
  magnitude of $Nu$ was affected by a normalization error, hence all
  points of Fig. (1) of that paper should be multiplied by a factor
  $240$ (corresponding to the grid size of our simulation). This of
  course does not affect the scaling exponent given there.}

\begin{figure}[htbp]
\begin{center}
\setlength{\unitlength}{1.0cm}
\begin{picture}(1,1)
\put(-3.5,-0.5){(a)}
\end{picture}
\includegraphics[width=8.cm]{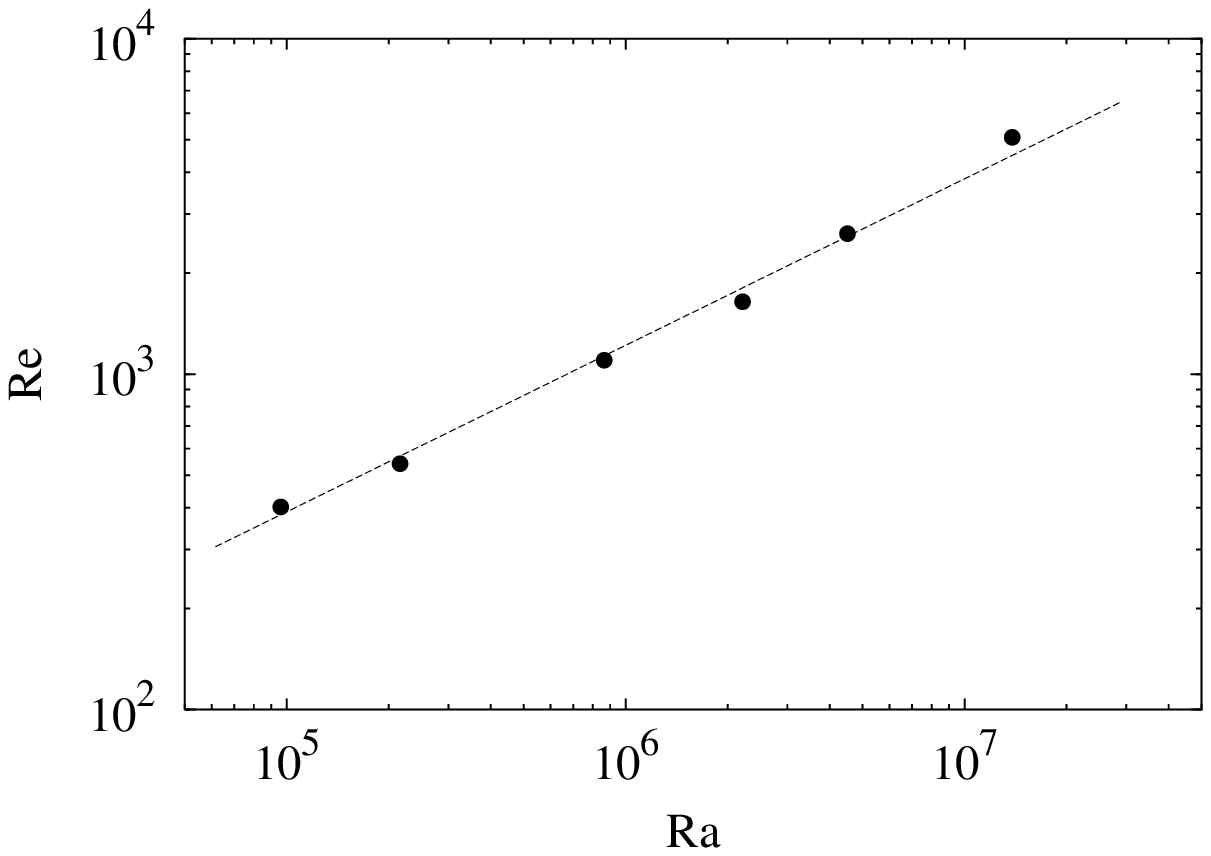}
\setlength{\unitlength}{1.0cm}
\begin{picture}(1,1)
\put(-3.5,-0.5){(b)}
\end{picture}
\includegraphics[width=8.cm]{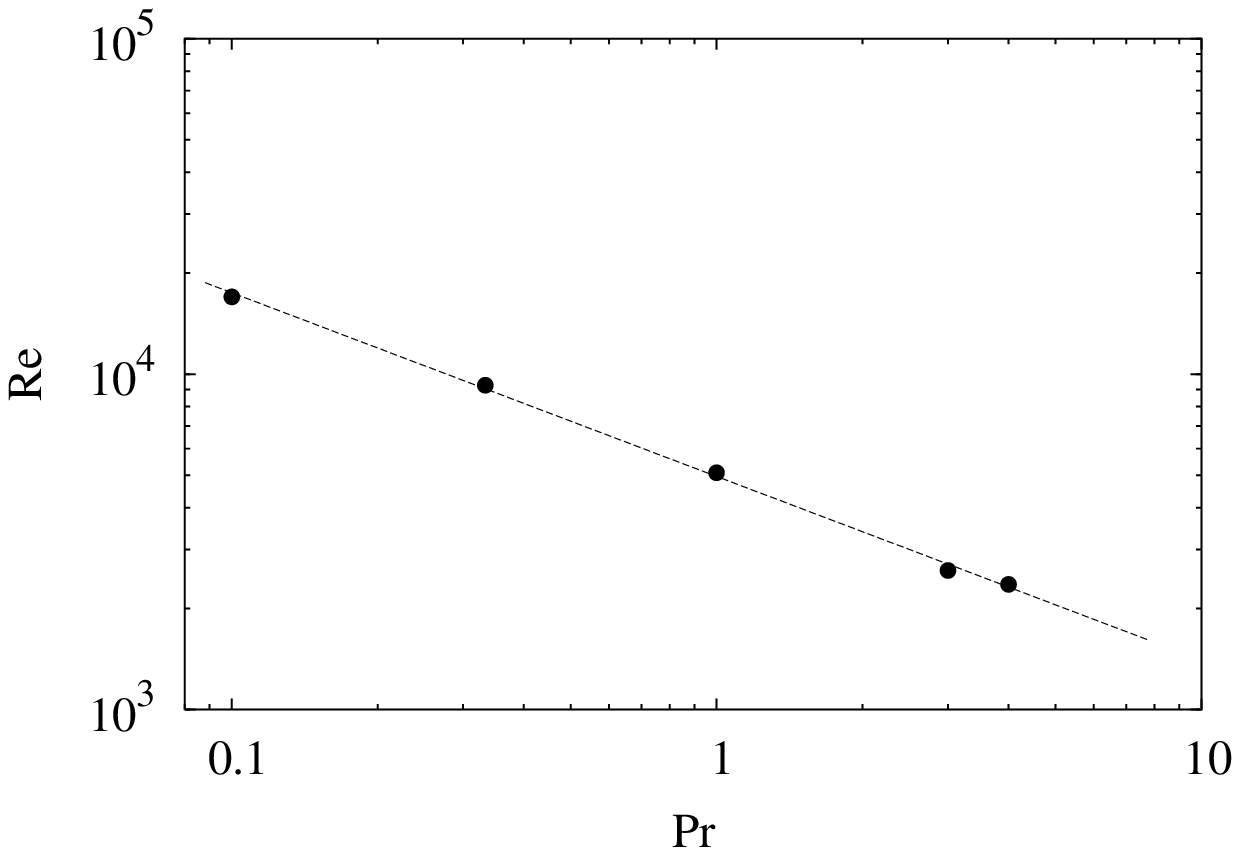}
\caption{
  (a) $Re(Ra)$ for $Pr = 1$, with a fitted slope $0.50 \pm 0.02$. (b)
  $Re(Pr)$ for $Ra= 1.4 \cdot 10^7$, with a fitted slope $-0.55 \pm
  0.01$.
\label{fig2}
}
\end{center}
\end{figure}

In Figure \ref{fig1}b we display $Nu$ as function of $Pr$ for fixed
$Ra=1.4\cdot 10^7$. For the cases with $Pr\ne 1$ the convergence of
the three different methods to calculate $Nu$ is not perfect. This may
be due to numerical errors in the resolution of the small scale
differences, especially when $\nu$ and $\kappa$ are considerably
different.  However, one can clearly notice a strong increase of $Nu$
with $Pr$. A fit with an effective power law gives $Nu \sim Pr^{0.43
  \pm 0.07}$, which is consistent with the asymptotic power law $Nu
\sim Pr^{1/2}$ suggested by the GL theory and by the small $Pr$ regime
(\ref{eq1}) suggested by Kraichnan, but not with Kraichnan's large
$Pr$ regime (\ref{eq2}). Increasing further $Pr$ (at fixed $Ra$) the
flow will eventually laminarize, i.e., can no longer be considered as
model system for the bulk of turbulence. This also follows from Figure
\ref{fig2}b, in which we show the Reynolds number: 
\be 
Re = {u^\prime L\over \nu} 
\label{eq10} 
\ee 

as function of $Pr$ for fixed $Ra= 1.4 \cdot 10^7$. Note that this is
the {\it fluctuation} Reynolds number, defined by the rms velocity
fluctuation $u'=\left<u^2\right>^{1/2}$: in homogeneous RB no large
scale wind exists.  $Re(Pr)$ displays an effective scaling law $Re\sim
Pr^{-0.55 \pm 0.01}$, consistent with the GL prediction $Pr^{-1/2}$
for the ultimate regime (if one identifies the wind Reynolds number in
GL with the fluctuation Reynolds number here) and also with the
Kraichnan prediction (\ref{eq1}). Also the Ra scaling of Re is
consistent with GL (and also with Kraichnan), $Re \sim Ra^{1/2}$, as
seen from Figure \ref{fig2}a and as already shown in Ref.\ 
\cite{loh03}.

\section{Scaling laws for $\eps_u$, $\eps_\theta$ and the temperature fluctuations}

\subsection{ Kinetic  and thermal dissipations}
The homogeneous RB turbulence offers the opportunity to numerically
test one of the basic assumptions of the GL theory, namely, that the
energy dissipation rate in the bulk scales like

\be 
\eps_{u,bulk} \sim {\nu^3\over L^4} Re^{3}.
\label{eq11}
\ee 

\begin{figure}[htbp]
\begin{center}
\setlength{\unitlength}{1.0cm}
\begin{picture}(1,1)
\put(-3.5,-0.5){(a)}
\end{picture}
\includegraphics[width=8.cm]{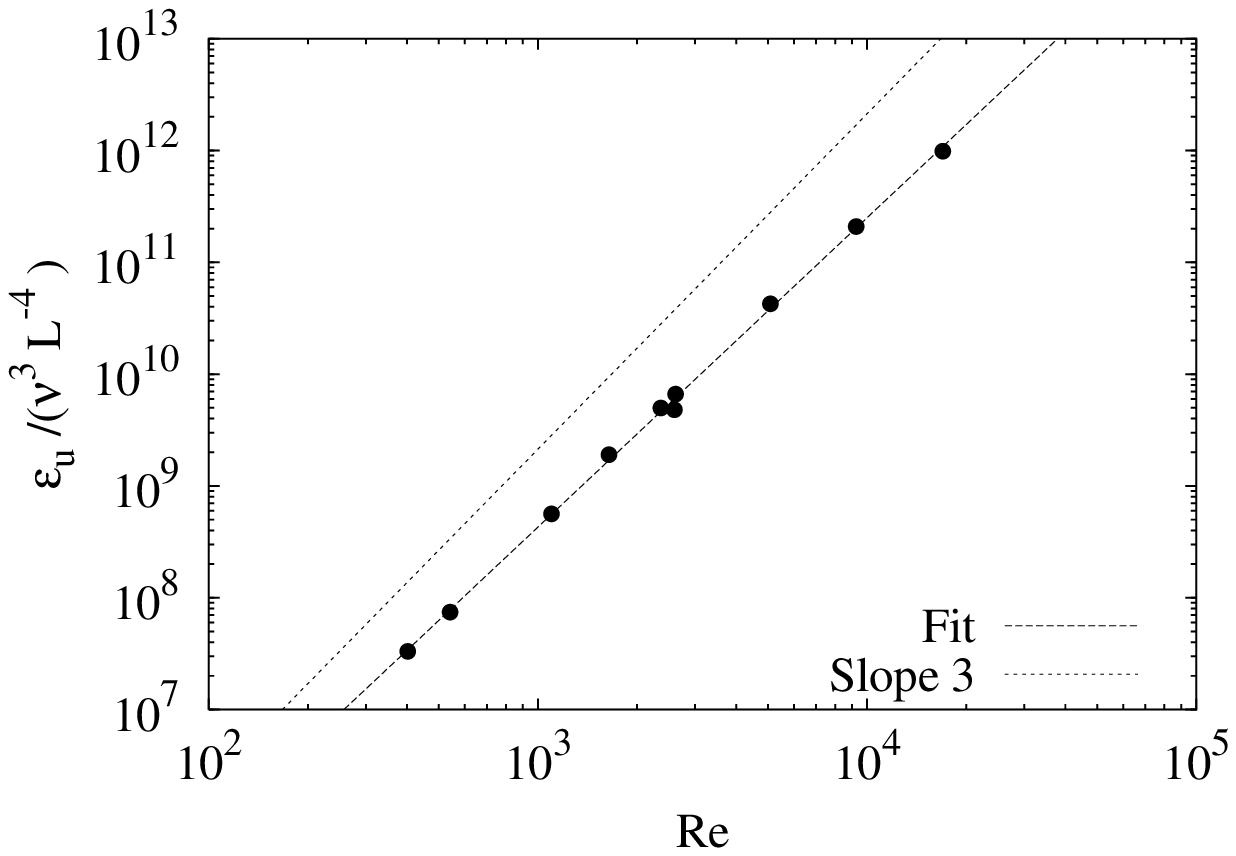}
\setlength{\unitlength}{1.0cm}
\begin{picture}(1,1)
\put(-3.5,-0.5){(b)}
\end{picture}
\includegraphics[width=8.cm]{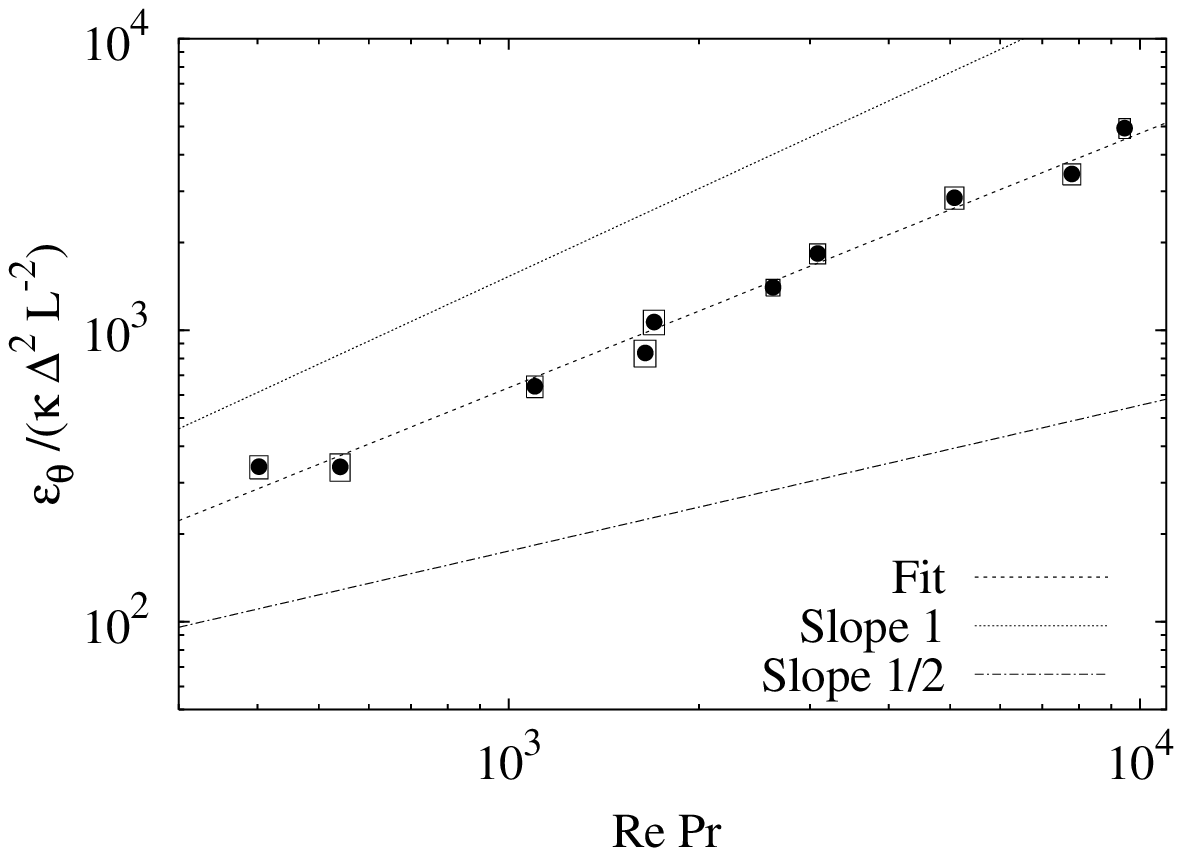}
\caption{
  (a) We show $\eps_u/(\nu^3 L^{-4})$ vs. $Re$. The fit gives a slope
  of $2.77 \pm 0.03$, slope $3$ is shown for comparison.  (b)
  $\eps_\theta/(\kappa \Delta^{2}L^{-2})$ vs. $RePr$.  We obtained a
  fitted slope $0.87 \pm 0.04$ while slopes $1$ and $1/2$ are also
  shown for comparison.
\label{fig3}
}
\end{center}
\end{figure}

In Figure \ref{fig3}(a) we plot $\eps_u/(\nu^3 L^{-4})$ vs. $Re$ for
all $Ra$ and $Pr$ and find $\eps_u/(\nu^3 L^{-4})\sim Re^{2.77 \pm
  0.03}$, close to the expectation (\ref{eq11}).

The disentanglement of the thermal dissipation rate $\eps_\theta$ into
two different scaling contributions is less straightforward. The GL
theory decombines: 
\be \eps_\theta = c_3 (RePr)^{1/2} + c_4 (RePr),
\label{eq12}
\ee 
where the first term has been interpreted as boundary layer and plume
contribution $\eps_{\theta , pl}$ and the second one as background
contribution $\eps_{\theta, bg}$ \cite{gro04}.  The prefactors $c_3$
and $c_4$ are given in Ref.\cite{gro01}.  Plumes are interpreted as
detached boundary layer \cite{gro04}. For homogeneous RB turbulence
one would expect the background contributions to be dominant as there
is no boundary layer.  But still some plumes may also develop in the
bulk and this is confirmed by the fact that we find a scaling law in
between the asymptotes $(RePr)^{1/2}$ and $RePr$, namely, $\eps_\theta
\sim(RePr)^{0.87\pm 0.04} $: closer to the background behavior just as
one would guessed.

\subsection{Temperature fluctuations}
\begin{figure}[htbp]
\begin{center}
\setlength{\unitlength}{1.0cm}
\begin{picture}(1,1)
\put(-3.5,-0.5){(a)}
\end{picture}
\includegraphics[width=8.cm]{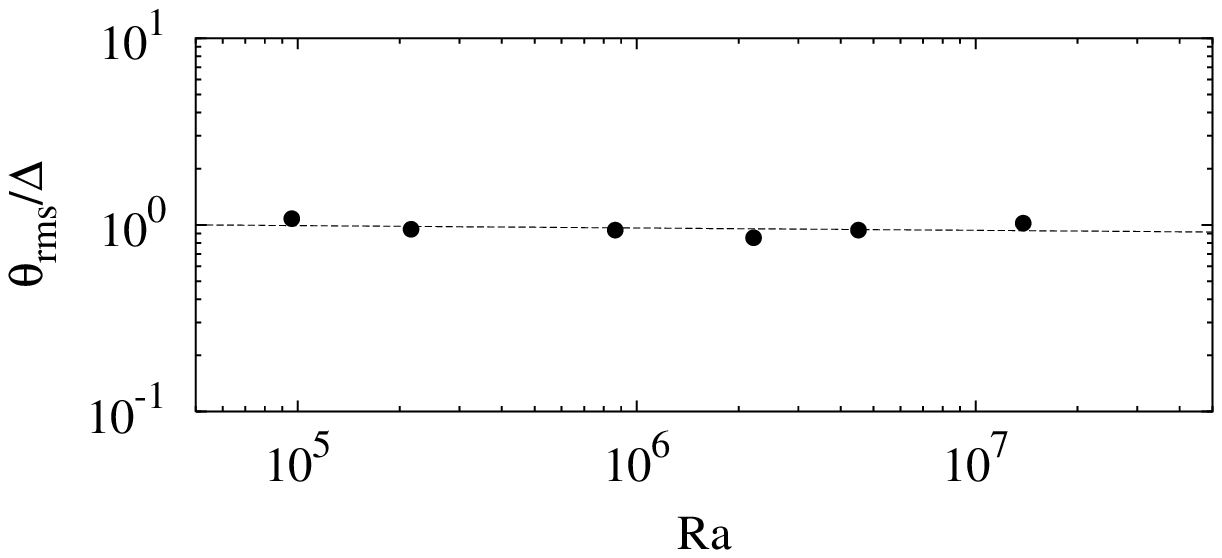}
\begin{picture}(1,1)
\put(-3.5,-0.5){(b)}
\end{picture}
\includegraphics[width=8.cm]{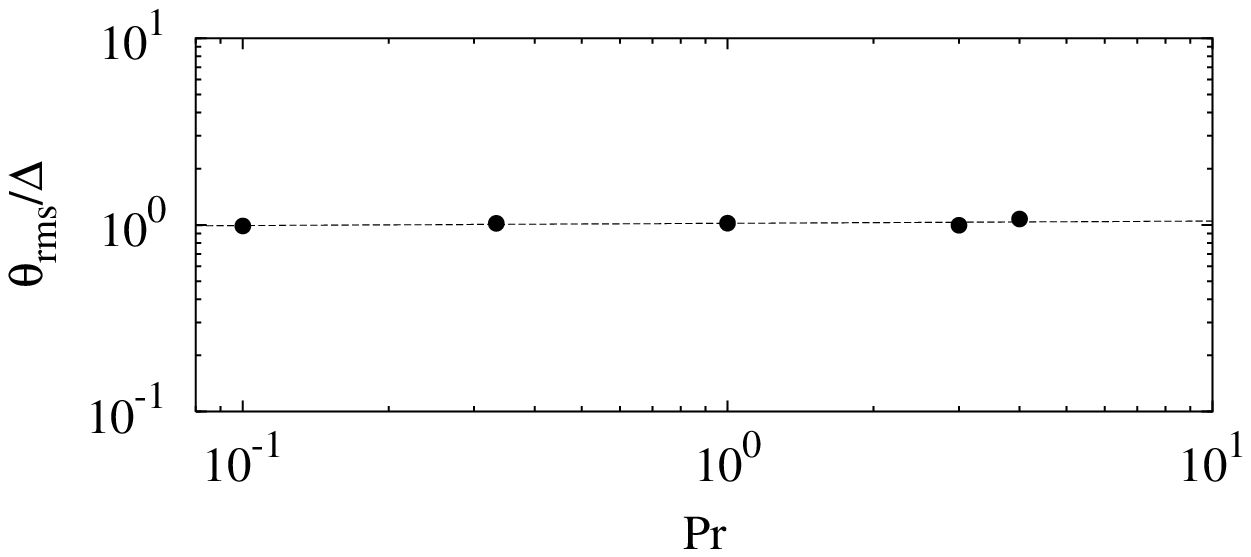}
\caption{
  (a) Normalized temperature variance $\theta'/\Delta$ vs. $Ra$ at
  fixed $Pr=1$.  (b) $\theta'/\Delta$ vs.  $Pr$ at fixed $Ra=1.4\cdot
  10^7$.
\label{fig4}
}
\end{center}
\end{figure}

In our numerics we find the temperature fluctuations $\theta' = \left<
  \theta^2 \right>^{1/2}$ to be independent from $Ra$ and $Pr$, see
Figure\ \ref{fig4}. That figures shows that we  have $\theta'
\simeq \Delta$ for all $Ra$ and $Pr$ within our numerical precision. In
contrast, Ref.\ \cite{gro04} predicted a dependence of the thermal
fluctuations on both Ra and Pr, namely $\theta'/\Delta \sim (Pr
Ra)^{-1/8}$ for the regimes IV$_l$ and IV$_l^\prime$ which correspond
to the bulk of turbulence analysed here.  Our interpretation of Figure
\ref{fig4} is that the bulk turbulence only has one temperature scale,
namely $\Delta$.  For real RB turbulence it is the boundary layer
dynamics which introduces further temperature scales, leading to the
$Ra$ and $Pr$ number dependence of the temperature fluctuations
observed in experiments \cite{nie00,cas89,day01,day02}.

\section{Dynamics of the flow}
\begin{figure}[!t]
\begin{center}
\includegraphics[width=\hsize]{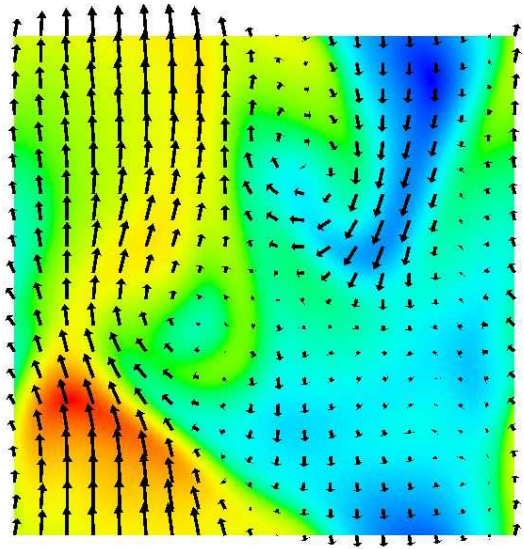}
\caption{
  Snapshot of the flow, showing elevator modes and jets.  Here
  $\theta$ is shown in colors: red and yellow encode for positive
  values, with red greater in amplitude than yellow, green is for
  small values around zero, while blue stands for the negative values,
  the dark blue stands for the more negative values. Velocity in the
  same plane is shown with arrows.
\label{fig5}
}
\end{center}
\end{figure}

In this section we provide an insight into the dynamics of the
periodic Rayleigh-B\'enard flow.  A bi-dimensional vertical snapshot
of the flow is shown in Figure \ref{fig5}. Already from this pictorial
view the presence of an upward moving hot column and a downward moving
cold column is clearly evident.

Indeed these large scale structure can be related to the presence of
``elevator modes'' (or jets, forming in the flow) growing in time
until finally breaking down due to some instability mechanisms.

%
%
As proposed by Doering and collaborators in \cite{doe05} it is
possible to predict the presence of these modes directly starting from
equations (\ref{eq5}) and (\ref{eq6}).  Doering {\it et al.} showed
that, due to the periodic boundary conditions, this coupled system of
equations admits a particular solution $\theta = \theta_0e^{\kappa
  \lambda t} \sin(\kk \cdot \xx)$, $u_3 = u_0e^{\kappa \lambda t}
\sin(\kk \cdot \xx)$, $u_2 = u_1 = 0$, which is independent from the
vertical coordinate $z$ (here $\kk=(k_x,k_y)$) and with:

\bea
\label{lambda}
\lambda &=& -{1\over 2}(Pr+1)k^2 + \\
\nonumber
&+& {1\over 2}\sqrt{(Pr+1)^2k^4 +4Pr\lp{Ra\over L^4}-k^4\rp}
\eea 

From equation (\ref{lambda}) one finds that the first unstable mode
appears for $Ra\ge Ra_c = (2\pi)^4 \sim 1558.54$, corresponding to the
instability of the smallest possible wavenumber in the system, i.e.
$k^2=({2\pi / L})^2 n^2$ with $n=(1,0)$.

\begin{figure}[htbp]
\begin{center}
\setlength{\unitlength}{1.0cm}
\begin{picture}(1,1)
\put(-3.5,-0.5){(a)}
\end{picture}
\includegraphics[width=8.cm]{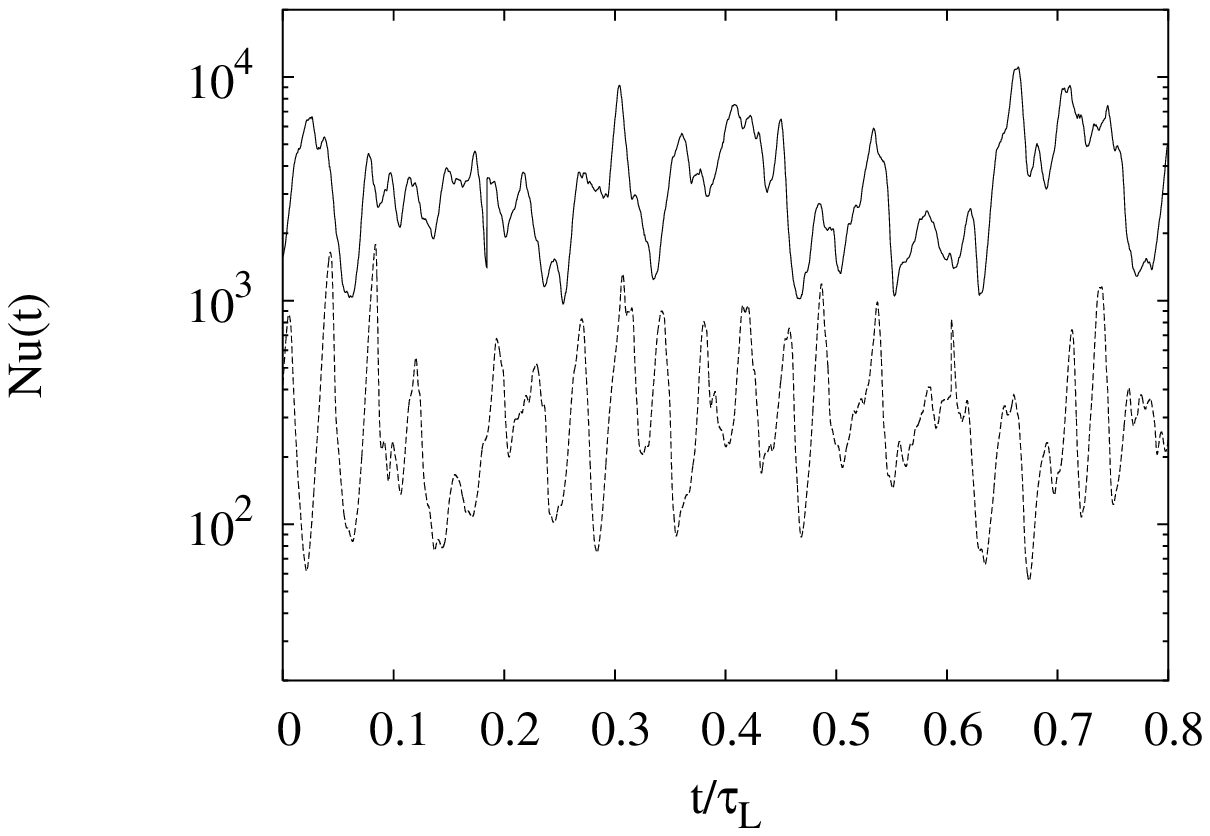}
\setlength{\unitlength}{1.0cm}
\begin{picture}(1,1)
\put(-3.5,-0.5){(b)}
\end{picture}
\includegraphics[width=8.cm]{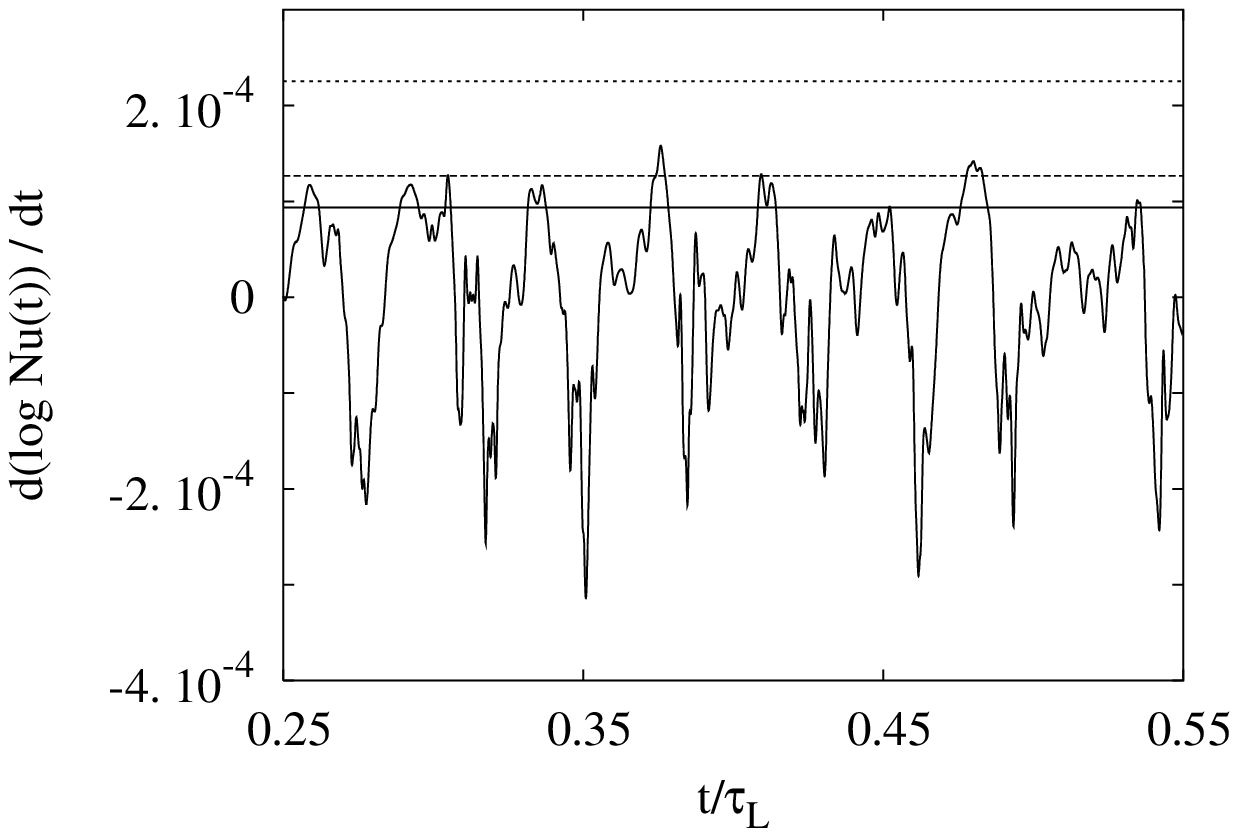}
\caption{
  (a) Time series $Nu(t)$ for $Ra = 1.4\cdot10^7$ (top) and
  $Ra=9.6\cdot10^4$ (bottom), in both case $Pr=1$.  (b) Logarithmic derivative of $Nu(t)$
  for $Ra=9.6\cdot10^4$, here reproduced only for a small time section
  of the data in (a). The series of horizontal lines represent the
  exponential rate of growing respectively (top to bottom) for the
  mode $\lambda(0,1)$, $\lambda(0,2)$, and $\lambda(1,2)$.
\label{fig6}
}
\end{center}
\end{figure}

The presence of accelerating modes with growth rate controlled by
$\lambda$ can also be seen from Figure \ref{fig6} where we show
$Nu(t)$ on log-scale (notice the huge range over which $Nu$
fluctuates).

\begin{figure}[htbp]
\begin{center}
\setlength{\unitlength}{1.0cm}
\begin{picture}(1,1)
\put(-3.5,-0.5){(a)}
\end{picture}
\includegraphics[width=8.cm]{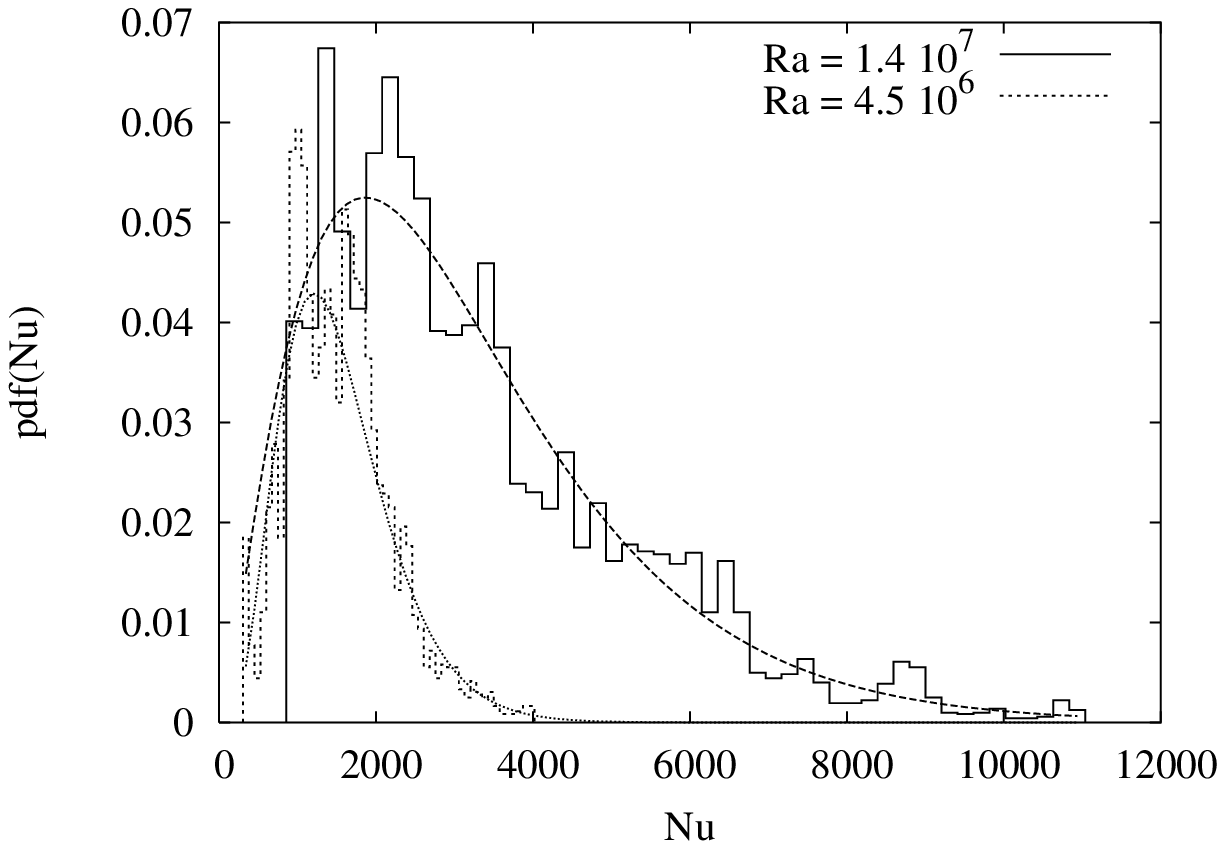}
\setlength{\unitlength}{1.0cm}
\begin{picture}(1,1)
\put(-3.5,-0.5){(b)}
\end{picture}
\includegraphics[width=8.cm]{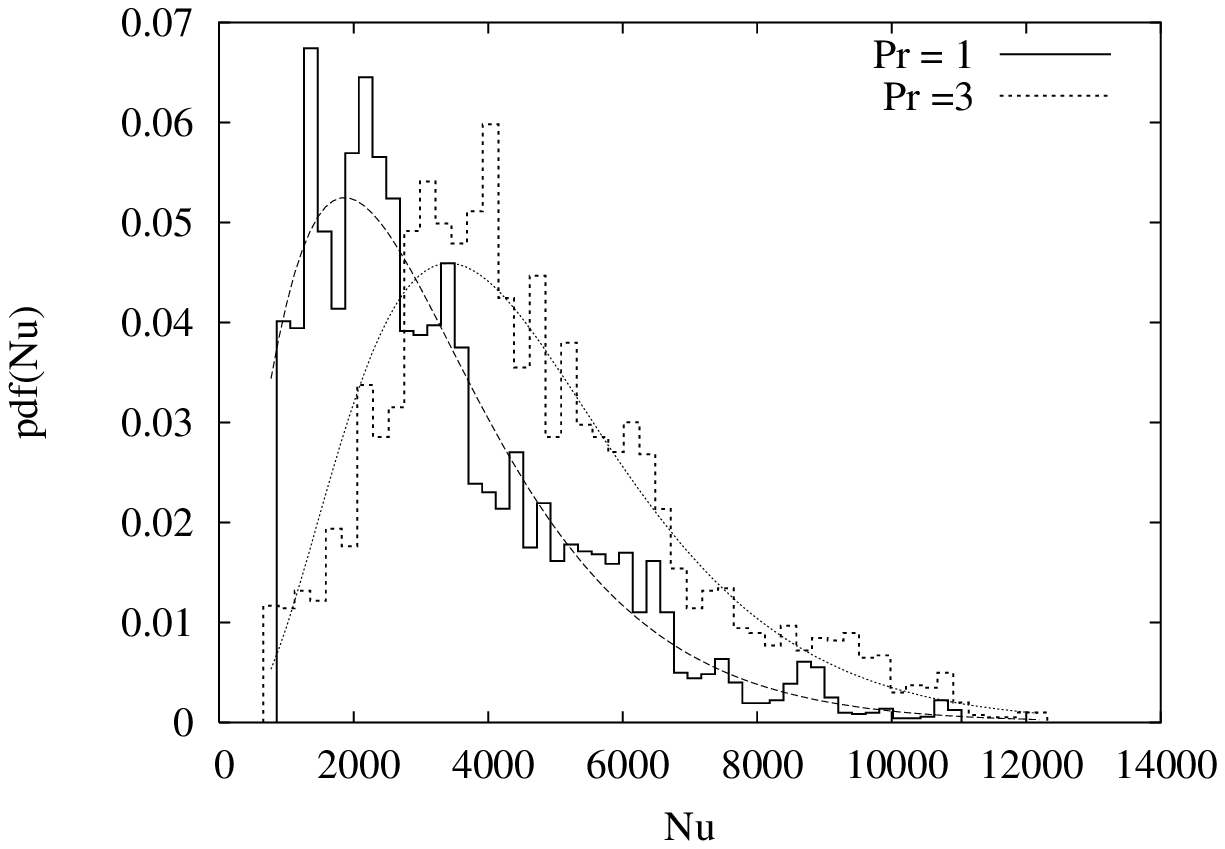}
\caption{
  (a) and (b): PDF of $Nu(t)$ for different $Ra$ and $Pr$. The
  superimposed curves correspond to a two parameter Gamma distribution
  fit, $Nu^a \exp( - b Nu)$ \cite{aum03}.
\label{fig7}
}
\end{center}
\end{figure}

\begin{figure}[!t]
\begin{center}
\setlength{\unitlength}{1.0cm}
\begin{picture}(1,1)
\put(-3.5,-0.5){(a)}
\end{picture}
\includegraphics[width=8.cm]{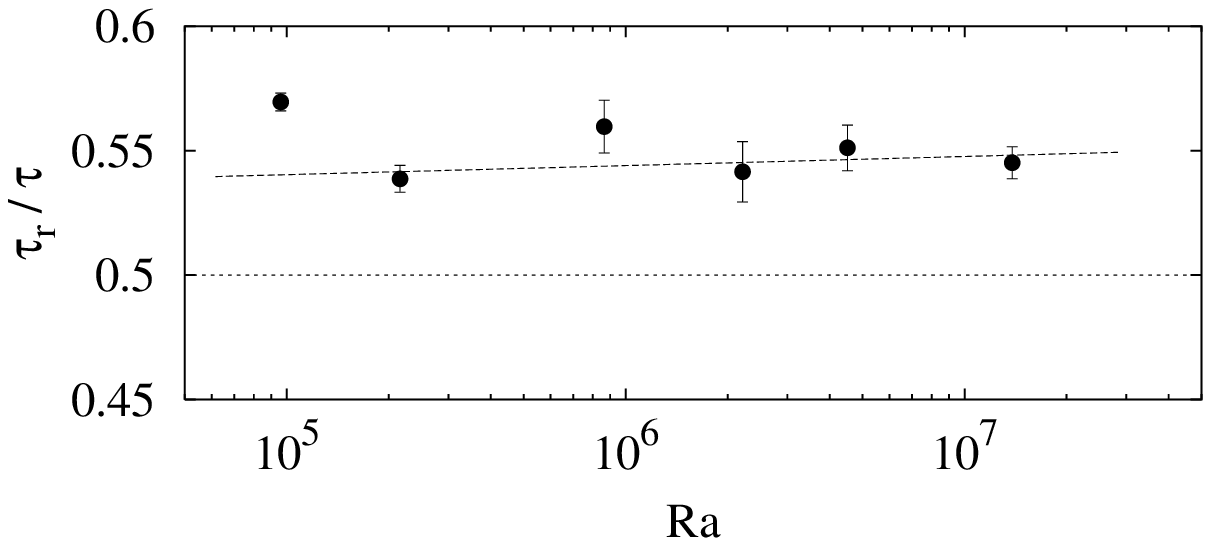}
\setlength{\unitlength}{1.0cm}
\begin{picture}(1,1)
\put(-3.5,-0.5){(b)}
\end{picture}
\includegraphics[width=8.cm]{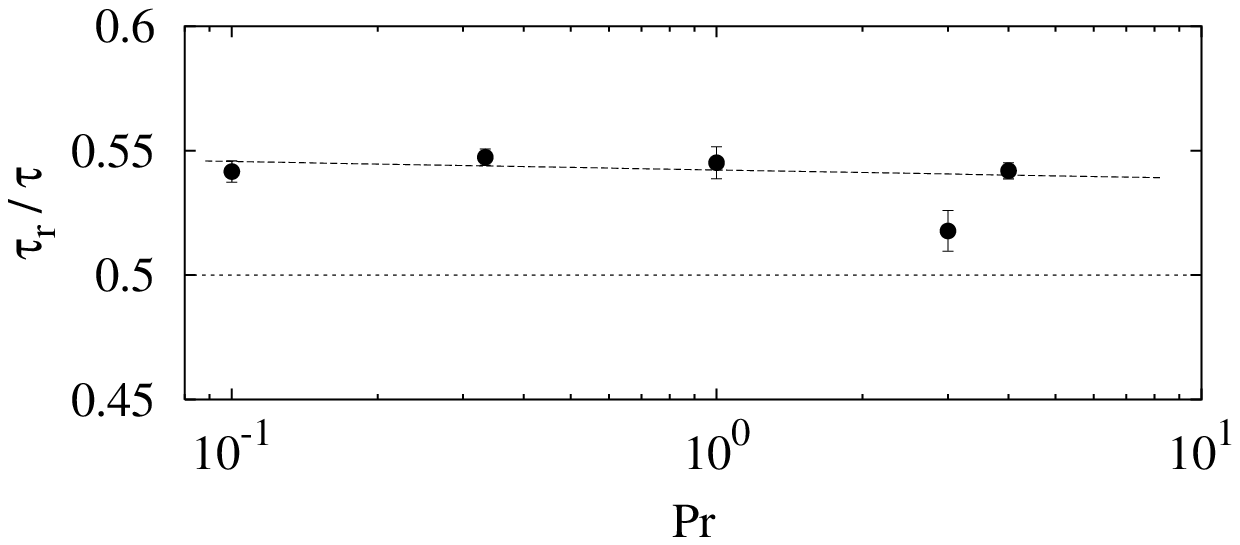}
\caption{
  Normalized rising time $\tau_r/\tau$ as a function of (a) $Ra$ for
  $Pr=1$ and (b) $Pr$ for $Ra = 1.4 \cdot 10^7$. The time $\tau_r$ is
  the total time with positive slope of $Nu(t)$, whereas the time
  $\tau$ is the total time of the run.  The slope of the two fits in
  the shown graphs is compatible with zero, the overall mean value for
  $\tau_r/\tau$ is $0.54$.
\label{fig9}
}
\end{center}
\end{figure}

In Figure \ref{fig7} we show the PDF of $Nu(t)$ which is strongly
skewed towards large $Nu$ values. This asymmetry reflects the periods
of exponential growth (also visible in Figure \ref{fig6}).  As can be
seen in Figure \ref{fig9}, for all $Ra$ and $Pr$ the system typically
spends 54\% of the time in growing modes.

\begin{figure}[htbp]
\begin{center}
\setlength{\unitlength}{1.0cm}
\begin{picture}(1,1)
\put(-3.5,-0.4){(a)}
\end{picture}
\includegraphics[width=8.cm]{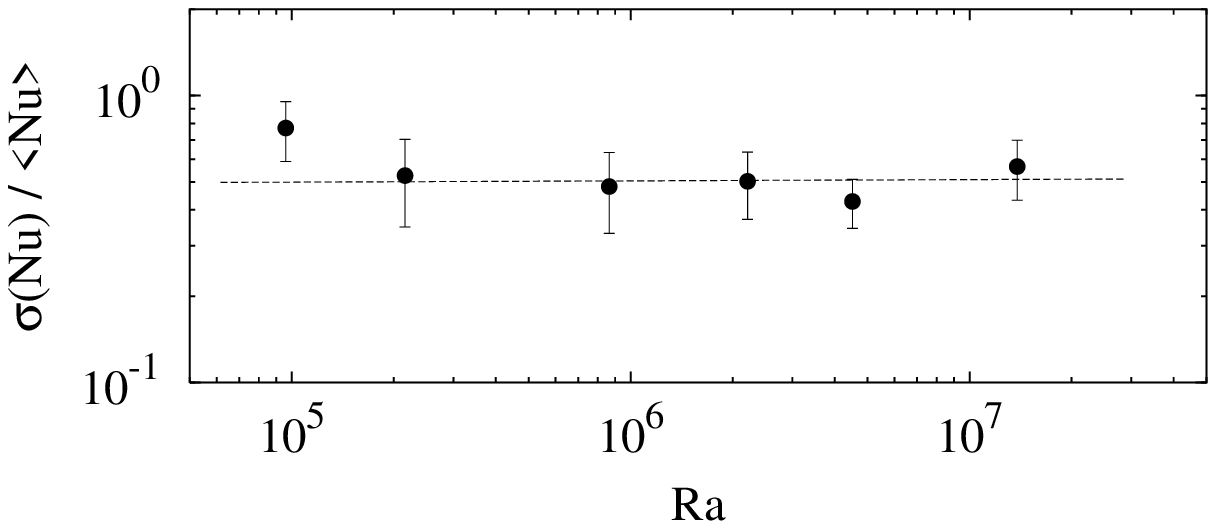}
\setlength{\unitlength}{1.0cm}
\begin{picture}(1,1)
\put(-3.5,-0.4){(b)}
\end{picture}
\includegraphics[width=8.cm]{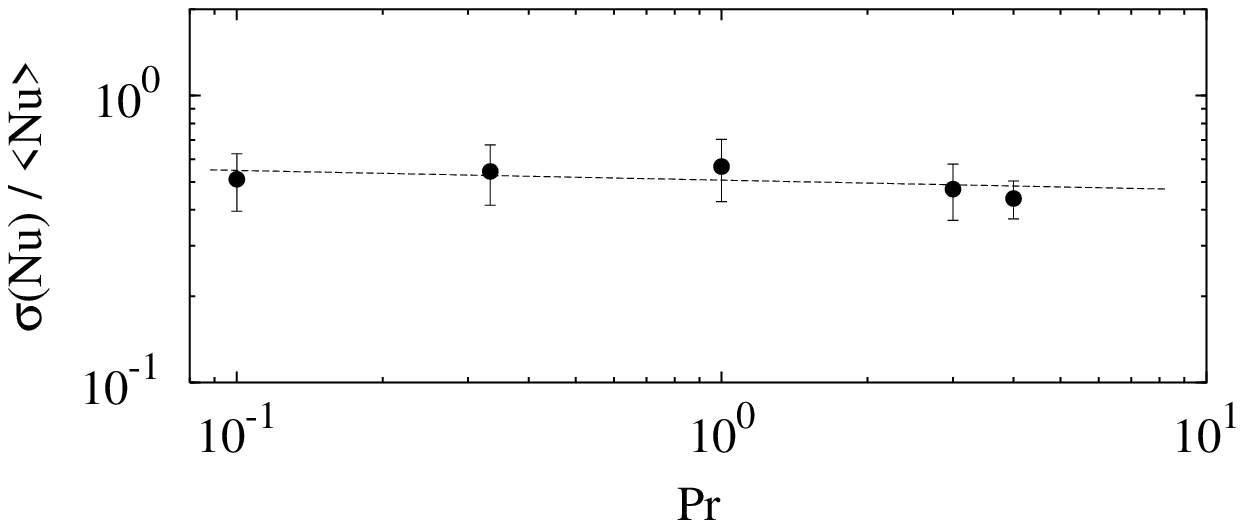}
\caption{
  Relative fluctuations $\sigma(Nu)/\left< Nu \right>$, where
  $\sigma(Nu)\equiv \left< ( Nu (t) - \left<Nu\right> )^2
  \right>^{1/2}$, as function of (a) $Ra$ for $Pr=1$ and (b) $Pr$ for
  $Ra = 1.4 \cdot 10^7$.
\label{fig8}
}
\end{center}
\end{figure}

Also the relative fluctuations of $Nu$ on the $Ra$ and $Pr$ numbers
(see Figure \ref{fig8}) seems to indicate no dependencies, at least in
the range of parameters studied.

Despite the presence of exact exploding solutions, our system clearly
shows that in the turbulent regime these solutions become unstable due
to some yet to be explored instability mechanism. The interplay
between exploding modes and destabilization sets the value of the
Nusselt number, i.e.  the heat transfer through the cell.

We stress that the study of the dynamics of the explosive solutions
and of their successive collaps in a turbulent cell is crucial for the
understanding the behaviour of ``integral'' quantities, like, for
example, the heat transfer.

\section{Conclusions}
In conclusion, we confirmed that both the $Ra$- and the $Pr$-scaling of $Nu$
and $Re$ in homogeneous Rayleigh-Benard convection is consistent with
the suggested scaling laws of the Grossmann-Lohse theory for the
bulk-dominated regime (regime IV$_l$ of \cite{gro00,gro01,gro02}),
which is the so-called ``ultimate regime of thermal convection''. We
also showed that the thermal and kinetic dissipations scale roughly as
assumed in that theory. The temperature fluctuations do not show any
$Ra$ or $Pr$ dependence for homogeneous Rayleigh-Benard convection. From
the dynamics the heat transport and flow visualizations we identify
``elevator modes'' which are brought into the context of a recent
analytical finding by Doering {\it et al.} In future work we plan to
further elucidate the flow organization and in particular the
instability mechanisms of the elevator modes which set the Nusselt
number in homogeneous RB flow and therefore presumably also in the
ultimate regime of thermal convection.

\begin{acknowledgments}
  We thank Charlie Doering for stimulating discussions on section V.
  D.L.\ wishes to thank Siegfried Grossmann for extensive discussions
  and exchange over the years.  This work is part of the research
  programme of the Stichting voor Fundamenteel Onderzoek der Materie
  (FOM), which is financially supported by the Nederlandse Organisatie
  voor Wetenschappelijk Onderzoek (NWO).  Support by the European
  Union under contract HPRN-CT-2000-00162 ``Non Ideal Turbulence'' is
  also acknowledged.  This research was also supported by the INFN,
  through access to the APE$mille$ computer resources. E.C. has been
  supported by Neuricam spa (Trento, Italy) in the framework of a
  doctoral grant program with the University of Ferrara and during his
  visit at University of Twente by SARA through the HPC-Europe
  program.
\end{acknowledgments}

\end{document}